\documentclass[aps,prd,twocolumn,superscriptaddress,showpacs,showkeys,nofootinbib]{revtex4-1}

\usepackage{latexsym}
\usepackage{amsmath}
\usepackage{amssymb}
\usepackage{amsfonts}

\usepackage{color}

\usepackage{supertabular} 
\usepackage{placeins}
\usepackage{epsfig}
\usepackage{graphicx}

\usepackage{soul} 

\begin{document}


\title{The strange partner of the  $Z_{c}$ structures in a coupled-channels model}

\author{Pablo G. Ortega}
\email[]{pgortega@usal.es}
\affiliation{Departamento de Física Fundamental and Instituto Universitario de F\'isica 
Fundamental y Matem\'aticas (IUFFyM), Universidad de Salamanca, E-37008 Salamanca, Spain}

\author{David R. Entem}
\email[]{entem@usal.es}
\affiliation{Grupo de F\'isica Nuclear and Instituto Universitario de F\'isica Fundamental y Matem\'aticas (IUFFyM), Universidad de Salamanca, E-37008 Salamanca, Spain}

\author{Francisco Fern\'andez}
\email[]{fdz@usal.es}
\affiliation{Grupo de F\'isica Nuclear and Instituto Universitario de F\'isica 
Fundamental y Matem\'aticas (IUFFyM), Universidad de Salamanca, E-37008 
Salamanca, Spain}


\date{\today}

\begin{abstract}

The discovery of a new charged structure in the $K^+$ recoil-mass spectrum near the $D^-_s D^{*0}/D^{*-}_sD^0$ threshold, dubbed $Z_{cs}(3985)^-$, reinforce the idea that the structure of hadrons goes beyond the naive $qqq$ and the $q\bar q$ structures. 

The existence of this state, with quark content $c\bar c s\bar u$, can be expected from the well-established $Z_c(3900)^\pm$ and $Z_c(4020)$ states using SU(3) flavor symmetry. The $Z_c$ structures have been explained using the chiral constituent quark model in a coupled-channels calculation and, in this work, we undertake the study of the $Z_{cs}(3985)^-$ using the same model.

We are able to reproduce the $K^+$ recoil-mass spectrum without any fine tuning of the model parameters. The study of the analytical structure of the S-matrix allows us to conclude that the structure is due to the presence of one virtual pole. A second state, the SU(3) flavor partner of the $Z_c(4020)$ is predicted at $\sim\!\! 4110$ MeV/$c^2$. New states in the hidden bottom strange sector are also predicted.

\end{abstract}



\maketitle


\section{Introduction} \label{sec:introduction}

Signals of non-conventional meson structures have appeared in the so-called B-factories and other accelerator facilities in the last years.
Among them, the more significant ones are the meson charged structures, called collectively the $Z$ structures, that suggest a minimum contain of four quarks to describe the state.
Going back in time, ten years ago the Belle Collaboration~\cite{Belle:2011aa} announced the discovery the $Z_b(10610)^\pm$ and the $Z_b(10650)^\pm$, a pair of charged hidden-bottom resonances with $I^G(J^{PC})=1^-(1^{+-})$ in the $\Upsilon(5S)\to \pi^+\pi^-\Upsilon(nS)$ reaction. They are close to the $B\bar B^\ast$ and $B^\ast \bar B^\ast$ thresholds, respectively. Later on, the BESIII and Belle Collaborations discovered another charged state, dubbed $Z_{c}(3900)^\pm$~\cite{Ablikim:2013mio}, in the $\pi^+\pi^- J/\psi$ invariant mass spectrum of the $e^+e^-\rightarrow\pi^+\pi^- J/\psi$ process at $\sqrt{s}=4.26$ GeV.  Its neutral partner was also reported in Ref.~\cite{Ablikim:2015tbp}.
Soon after all these experimental activities, the BESIII Collaboration reported the discovery of another charged state, the $Z_{c}(4020)^\pm$ resonance, in the $e^+e^-\to \pi^+\pi^- h_c$ channel with a mass of $M=(4022.9\pm 0.8\pm 2.7)$ MeV/$c^2$ and a width of $\Gamma=(7.9\pm 2.7\pm 2.6)$ MeV~\cite{Ablikim:2013wzq}. Its neutral partner was found by the BESIII Collaboration in Ref.~\cite{Ablikim:2014dxl}.

More recently, the BESIII Collaboration has observed a new signal, the $Z_{cs}(3985)^-$, in the $K^+$ recoil-mass spectrum in the process $e^+e^-\rightarrow K^+(D_s^-D^{*0}+D_s^{*-}D^0)$ at the center of mass energy $\sqrt{s}=4.681$ GeV~\cite{Ablikim:2020hsk}. Its pole mass and width were determined with a mass dependent width Breit-Wigner line shape as

\begin{align}
M^{pole}_{Z_{cs}}&=(3982.5^{+1.8}_{-2.6}\pm 2.1) {\rm MeV}/c^2, \\
\Gamma^{pole}_{z_{cs}}&= (12.8^{+5.3}_{-4.4}\pm 3.0) {\rm MeV},
\end{align}
being the first uncertainties statistical and the second ones systematic.

From the experimental analysis~\cite{Ablikim:2020hsk}, the minimum quark contain of the $Z_{cs}(3985)^-$ is most likely $c\bar c s\bar u$. This new state has two remarkable characteristics: It is close to the $D_s^-D^{*0}$ and $D_s^{*-}D^0$ thresholds and its mass is about $100$ MeV/$c^2$ larger than that of the $Z_c(3900)^\pm$, which is the typical mass difference between $D^{(*)}_s$ and $D^{(*)}$ mesons. These two characteristics suggest, first, that the $Z_{cs}(3985)^-$ may have a significant molecular $D_s^-D^{*0}+D_s^{*-}D^0$ component and, second, that it could be the $SU(3)$ flavor partner of the $Z_c(3900)^\pm$, where a $u$ or $d$ quark is replaced by a $s$ quark in its quark content.

This possibility has been explored in Refs~\cite{Yang:2020nrt,PhysRevD.102.111502}. In Ref.~\cite{Yang:2020nrt}, using a toy model,  although a more complete calculation is also performed in the same reference, it is shown that the mass of the $Z_{cs}$ and a possible $Z^*_{cs}$ can be deduced from the $Z_c(3900)^\pm$ and the $Z_c^*(4020)^\pm$ structures by applying $SU(3)_F$ symmetry. A similar conclusion can be found in Ref.~\cite{PhysRevD.102.111502}.

Shortly after the discovery of the $Z_{cs}(3985)^-$, several theoretical works have proposed different explanations for its structure, including QCD sum rules~\cite{Wan:2020oxt,Wang:2020rcx,Wang:2020iqt,Azizi:2020zyq,Xu:2020evn,Wang:2020rcx,Albuquerque:2021tqd,Ozdem:2021yvo}, one boson exchange models~\cite{Chen:2020yvq}, quarks models~\cite{Jin:2020yjn} and effective field theories~\cite{Yang:2020nrt,Du:2020vwb,PhysRevD.102.111502}. Most of the calculations reproduced the mass and, in some cases, the width of the $Z_{cs}(3985)^-$ as a $D_s^-D^{*0}+D_s^{*-}D^0$ $I(J^{P})=\frac{1}{2}(1^{+})$ molecular state.

Besides its mass, it is interesting to look at the line shapes of the $K^+$ recoil-mass spectrum, because either the Breit-Wigner parametrization used by the experimental analysis could not correspond to the S-matrix physical poles or a too-weak interaction cannot be enough to generate a bound states or resonance, but can generate a cusp structure in the $D_s^-D^{*0}+D_s^{*-}D^0$ thresholds sufficient to describe the experimental data. Fits to the line shapes can be found in Refs.~\cite{Yang:2020nrt,IKENO2021136120,Wang:2020htx}.

The LHCb Collaboration, very recently, has also announced the observation of $c\bar c u\bar s$ exotic states decaying to $J/\psi K^+$ final channels, in the reaction $B^+\to J/\psi\phi K^+$~\cite{Aaij:2021ivw}. They have reported the discovery of two $Z_{cs}$ $J^P=1^+$ states: The so-called $Z_{cs}(4000)^+$ resonance, with a mass of $4003\pm 6^{+4}_{-14}$ MeV/$c^2$ and a width of $131\pm15\pm26$ MeV, and the $Z_{cs}(4220)^+$, with a mass of $4216\pm 24^{+43}_{-30}$ MeV/$c^2$ and a width of $233\pm52^{+97}_{-73}$ MeV. Based on their experimental analysis, the authors find no evidence that the $Z_{cs}(4000)^+$ is the same as the $Z_{cs}(3985)^-$ observed by BESIII, though their energy proximity suggest a relation between both structures which is worth exploring. Indeed, in a recent added note, Ref.~\cite{Yang:2020nrt} points to the possibility that both structures are the same, but their masses and widths may not be consistent due to experimental resolution and/or coupled-channels effects.

In this work we investigate the structure of the $Z_{cs}(3985)^-$ to test the hypothesis that it is the $SU(3)_F$ partner of the $Z_c(3900)^\pm$ and analyze in what extent this symmetry is fulfilled, because the predictions of symmetries can be slightly modified depending on the relative position of the thresholds and the coupling between the different channels involved~\cite{Entem:2016ojz}. 

The $Z_c(3900)^\pm$ structure has been analyzed in a coupled-channels scheme based on the chiral constituent quark model~\cite{Ortega:2018cnm} (see Refs.~\cite{Valcarce:2005em,Segovia:2013wma} for a review of the model). The calculation 
was done for the $I^G(J^{PC})=1^+(1^{+-})$ sector and included the $\pi J/\psi$ (3234.19 MeV/$c^2$), $\rho\eta_c$ (3755.79 MeV/$c^2$), $D\bar D^\ast$ (3875.85 MeV/$c^2$), $D^\ast \bar D^\ast$ (4017.24 MeV/$c^2$) channels, that takes into account the most relevant decays and closest thresholds to the experimental masses of the $Z_c(3900)^\pm$ and $Z_c(4020)$ hadrons, whose masses are shown in parenthesis.
The main conclusion of the calculation of Ref.~\cite{Ortega:2018cnm} is that the line shapes of the $D\bar D^*$, $\pi J/\psi$ and $D^*\bar D^*$ invariant-mass distributions are well reproduced without any fine-tuning of the model parameters. The peculiar characteristic of the result is that the diagonal interaction $\bar D^{*}D^{*}$ is suppressed in the $I=1$ state, being the non-diagonal interaction, due to the coupling with the other channels, responsible for the structures appeared in the line shapes. Then, a coupled-channels calculation is mandatory. Following this idea, in order to describe the $Z_{cs}(3985)^-$ we use the same chiral constituent quark model (CQM) including the following channels: $D^{*0}D_s^-$, $D^0 D_s^{*-}$, $D^{*0} D_s^{*-}$, $J/\psi K^{*\,-}(892)$, $\eta_c K^{*\,-}(892)$ and $J/\psi K^-$.

The structure of the present manuscript is organized in the following way: In Sec.~\ref{sec:theory} the theoretical framework is briefly presented and discussed, Sec.~\ref{sec:results} is devoted to the analysis and discussion on the obtained results and we summarize and give some conclusions in Sec.~\ref{sec:summary}.

\section{Theoretical Formalism}
\label{sec:theory}

\subsection{Chiral constituent quark model}

Chiral symmetry is one of the cornerstone of the Hadron Physics. The QCD lagrangian with massless quark is invariant under chiral rotations. However, this symmetry does not appear in nature (e.g. the mass splitting between the $\rho$ meson and its chiral partner the $a_1$ meson is about $300$ MeV/$c^2$). As a consecuence, a dynamical momentum dependent quark mass $M=M(q^2)$ and $M(q^2\to\infty)=m_q$ is developed, and a Goldstone-boson exchange interaction emerges between the light quarks. Let us precise what we understand by light quarks. Historically, the successful prediction of the $\Omega^-$ baryon taught us the importance of the $SU(3)_F$ symmetry, which means that the $u$, $d$ and $s$ quarks are treated as light quarks. Obviously, $SU(3)_F$ symmetry is not exact and we expect that it can be violated in some extent, but these effects can be hidden in the model parametrization.

The Lagrangian of our model tries to mimic the previous phenomena based on the following effective Lagrangian at low-energy~\cite{Diakonov:2002fq}
\begin{equation}
{\mathcal L} = \bar{\psi}(i\, {\slash\!\!\! \partial} -M(q^{2})U^{\gamma_{5}})\,\psi  \,,
\end{equation}
being $U^{\gamma_5} = e^{i\lambda _{a}\phi ^{a}\gamma _{5}/f_{\pi}}$ the matrix of Goldstone-boson fields, where $f_\pi$ is the pion decay constant, $\lambda^a$ are the $SU(3)$ colour matrices and $\phi^a$ denotes the pseudoscalar fields ($\vec \pi$,$K_i$,$\eta_8$), with $i=1,\ldots,4$.

This matrix of Goldstone-boson fields can be expanded as
\begin{equation}
U^{\gamma _{5}} = 1 + \frac{i}{f_{\pi}} \gamma^{5} \lambda_{a} \phi^{a} - \frac{1}{2f_{\pi}^{2}} \phi_{a} \phi^{a} + \ldots
\end{equation}
The constituent quark mass is obtained from the first term, the second one describes the pseudoscalar meson-exchange interaction among quarks and the main contribution of the third term comes from the two-pion exchange, which is modeled by means of a scalar-meson exchange potential.

Beyond the chiral-symmetry breaking scale, QCD perturbative effects appears. We take them into account through the one-gluon exchange potential derived from the following vertex Lagrangian term
\begin{equation}
{\mathcal L}_{qqg} = i\sqrt{4\pi\alpha_{s}} \, \bar{\psi} \gamma_{\mu} 
G^{\mu}_a \lambda^a \psi,\label{Lqqg}
\end{equation}
where $\alpha_{s}$ is the strong coupling constant and $G^{\mu}_a$ is the gluon field. The strong coupling constant, $\alpha_{s}$, has a scale dependence which allows a consistent description of light, strange and heavy mesons. Its explicit expression can be found, e.g., in Ref.~\cite{Vijande:2004he}.

Multigluon exchanges between quarks, which are supposed to be responsible of avoiding colored hadrons, are implemented phenomenologically as a confining interaction. In our CQM, the confinement is represented as a linear potential, due to multi-gluon exchanges between quarks, that is screened at large inter-quark distances as a consequence of sea quarks~\cite{Bali:2005fu}:
\begin{equation}
V_{\rm CON}(\vec{r}\,)=\left[-a_{c}(1-e^{-\mu_{c}r})+\Delta \right]  (\vec{\lambda}_{q}^{c}\cdot\vec{\lambda}_{\bar{q}}^{c}) \,.
\label{eq:conf}
\end{equation}
Here, $a_{c}$ and $\mu_{c}$ are model parameters. One can see that the potential is linear at short inter-quark distances with an effective confinement strength $\sigma = -a_{c} \, \mu_{c} \, (\vec{\lambda}^{c}_{i}\cdot \vec{\lambda}^{c}_{j})$, while it becomes constant at large distances. 

\begin{figure}[th]
 \includegraphics[width=\columnwidth]{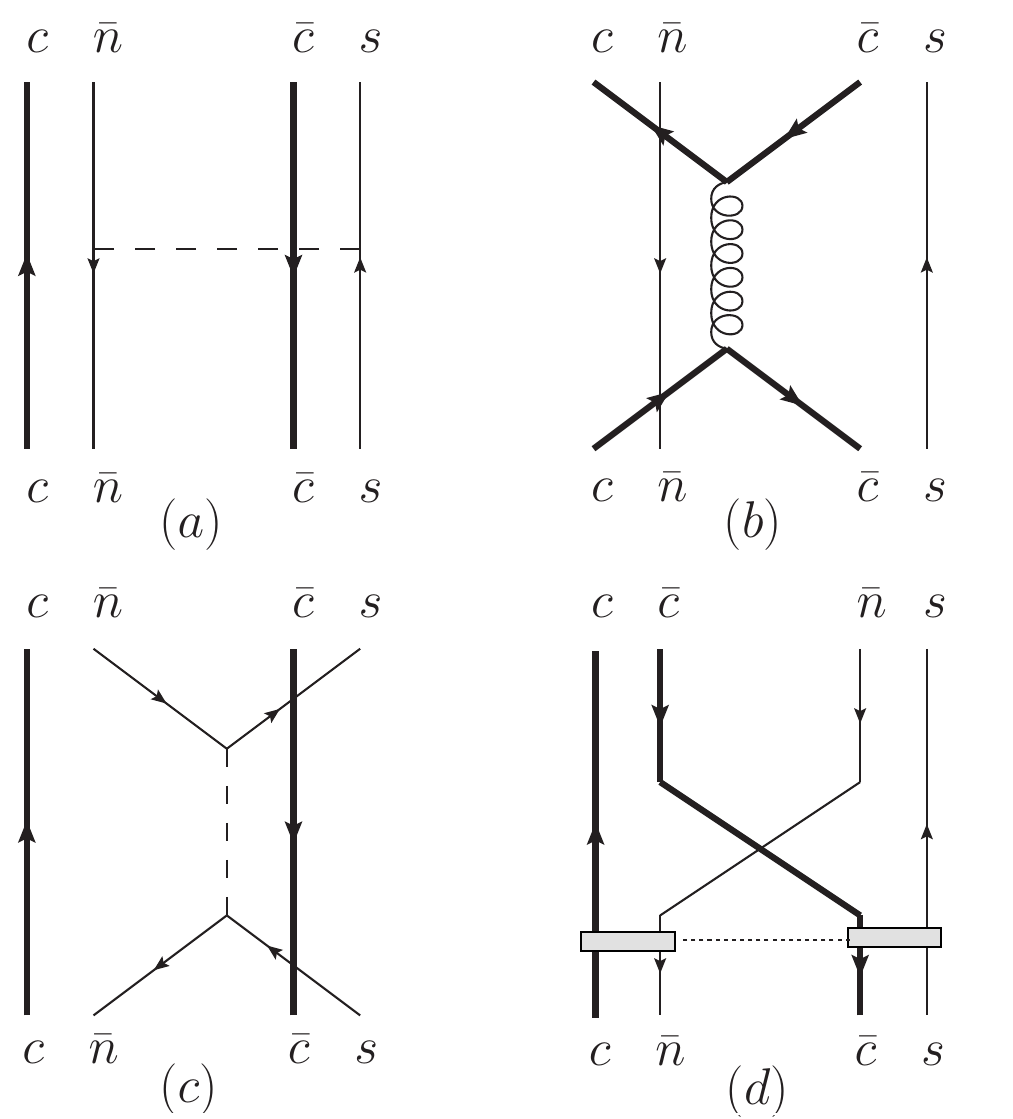} 
 \caption{\label{fig0} Different type of interactions considered in this work: a) Direct exchange of Goldstone bosons, b) annihilation diagram through a gluon, c) annihilation diagram through a kaon and d) quark rearrangement diagrams, where the gray band represents the sum of interactions between quarks of different clusters (see Eq.~\eqref{eq:Kernel}). }
\end{figure}

Besides the direct interaction, quarks and antiquarks can interact through annihilation processes. In the case of the $c\bar c s\bar u$ quark content, the annihilation proceeds through the one-gluon and one-kaon exchanges for the $c\bar c$ and $s\bar u$ pairs, respectively (see Fig.~\ref{fig0}).

A detailed physical background of the quark model can be found in Refs.~\cite{Vijande:2004he, Segovia:2008zz, Ortega:2011zza}. The model parameters and explicit expressions for the potentials can be also found therein. We want to highlight here that the interaction terms between light-light, light-heavy and heavy-heavy quarks are not the same in our formalism, i.e. while Goldstone-boson exchanges are considered when the two quarks are light, they do not appear in the other two configurations: light-heavy and heavy-heavy. However, the one-gluon exchange and confining potentials are flavor-blind.

\subsection{Resonating Group Method and Lippmann-Schwinger equation}

The aforementioned CQM specifies the microscopic interaction among constituent quarks. In order to describe the interaction at the meson level, we employ the Resonating Group Method~\cite{Wheeler:1937zza}, where mesons are considered as quark-antiquark clusters and an effective cluster-cluster interaction emerges from the underlying $q\bar q$ dynamics.

We assume that the wave function of a system composed of two mesons $A$ and $B$ with distinguishable quarks can be written as~\footnote{Note that, for simplicity of the discussion presented herein, we have dropped off the spin-isospin wave function, the product of the two color singlets and the wave function that describes the center-of-mass 
motion.}
\begin{equation}
\langle \vec{p}_{A} \vec{p}_{B} \vec{P} \vec{P}_{\rm c.m.} | \psi 
\rangle = \phi_{A}(\vec{p}_{A}) \phi_{B}(\vec{p}_{B}) 
\chi_{\alpha}(\vec{P}) \,,
\label{eq:wf}
\end{equation}
where  $\phi_{C}(\vec{p}_{C})$ is the wave function of a general meson $C$ with $\vec{p}_{C}$ the relative momentum between the quark and antiquark of the meson $C$. The wave function which takes into account the relative motion of the two mesons is $\chi_\alpha(\vec{P})$.

The projected Schr\"odinger equation for the relative wave function can be written as follows:
\begin{align}
&
\left(\frac{\vec{P}^{\prime 2}}{2\mu}-E \right) \chi_\alpha(\vec{P}') + \sum_{\alpha'}\int \Bigg[ {}^{\rm RGM}V_{D}^{\alpha\alpha'}(\vec{P}',\vec{P}_{i}) + \nonumber \\
&
+ {}^{\rm RGM}V_{R}^{\alpha\alpha'}(\vec{P}',\vec{P}_{i}) \Bigg] \chi_{\alpha'}(\vec{P}_{i})\, d\vec{P}_{i} = 0 \,,
\label{eq:Schrodinger}
\end{align}
where $E$ is the energy of the system. The direct potential ${}^{\rm RGM}V_{D}^{\alpha\alpha '}(\vec{P}',\vec{P}_{i})$ can be written as
\begin{align}
&
{}^{\rm RGM}V_{D}^{\alpha\alpha '}(\vec{P}',\vec{P}_{i}) = \sum_{i\in A, j\in B} \int d\vec{p}_{A'} d\vec{p}_{B'} d\vec{p}_{A} d\vec{p}_{B} \times \nonumber \\
&
\times \phi_{A}^{\ast}(\vec{p}_{A'}) \phi_{B}^{\ast}(\vec{p}_{B'}) 
V_{ij}^{\alpha\alpha '}(\vec{P}',\vec{P}_{i}) \phi_{A'}(\vec{p}_{A}) \phi_{B'}(\vec{p}_{B})  \,.
\end{align}
The quark rearrangement potential ${}^{\rm RGM}V_{R}^{\alpha\alpha'}(\vec{P}',\vec{P}_{i})$ represents a natural way to connect meson-meson channels with different quark content, such as $ J/\psi K^-$ and $D^{*\,0}D_s^-$ (see Fig.~\ref{fig0}(d)), and it is given by
\begin{align}
&
{}^{\rm RGM}V_{R}^{\alpha\alpha'}(\vec{P}',\vec{P}_{i}) = \sum_{i\in A, j\in B} \int d\vec{p}_{A'} 
d\vec{p}_{B'} d\vec{p}_{A} d\vec{p}_{B} d\vec{P} \phi_{A}^{\ast}(\vec{p}_{A'}) \times \nonumber \\
&
\times  \phi_{B}^{\ast}(\vec{p}_{B'}) 
V_{ij}^{\alpha\alpha '}(\vec{P}',\vec{P}) P_{mn} \left[\phi_{A'}(\vec{p}_{A}) \phi_{B'}(\vec{p}_{B}) \delta^{(3)}(\vec{P}-\vec{P}_{i}) \right] \,,
\label{eq:Kernel}
\end{align}
where $P_{mn}$ is the operator that exchanges quarks between clusters. 

The meson eigenstates $\phi_C(\vec{p}_{C})$ are calculated by means of the two-body
Schr\"odinger equation, using the Gaussian Expansion Method~\cite{Hiyama:2003cu}. This method provides enough accuracy and simplifies the subsequent evaluation of the needed matrix elements. With the aim of optimizing the Gaussian ranges employing a reduced number of free parameters, we use Gaussian trial functions whose ranges are given by a geometrical progression~\cite{Hiyama:2003cu}. This choice produces a dense distribution at short distances enabling better description of the dynamics mediated by short range potentials. 

The solution of the coupled-channels RGM equations is performed deriving from Eq.~\eqref{eq:Schrodinger} a set of coupled Lippmann-Schwinger equations of the form
\begin{align}
T_{\alpha}^{\alpha'}(E;p',p) &= V_{\alpha}^{\alpha'}(p',p) + \sum_{\alpha''} \int
dp''\, p^{\prime\prime2}\, V_{\alpha''}^{\alpha'}(p',p'') \nonumber \\
&
\times \frac{1}{E-{\cal E}_{\alpha''}(p^{''})}\, T_{\alpha}^{\alpha''}(E;p'',p) \,,
\end{align}
where $\alpha$ labels the set of quantum numbers needed to uniquely define a certain partial wave, $V_{\alpha}^{\alpha'}(p',p)$ is the projected potential that contains the direct and rearrangement potentials, and ${\cal E}_{\alpha''}(p'')$ is the energy corresponding to a momentum $p''$, written in the nonrelativistic case as:
\begin{equation}
{\cal E}_{\alpha}(p) = \frac{p^2}{2\mu_{\alpha}} + \Delta M_{\alpha} \,.
\end{equation}

Here, $\mu_{\alpha}$ is the reduced mass of the $AB$ system corresponding to the channel $\alpha$, and $\Delta M_{\alpha}$ is the difference between the threshold of the $AB$ system and the one we take as a reference.

We solve the coupled-channels Lippmann-Schwinger equation using the matrix-inversion method proposed in Ref.~\cite{Machleidt:1003bo}, generalized in order to include channels with different thresholds. Once the $T$-matrix is calculated, we determine the on-shell part which is directly related to the scattering matrix (in the case of nonrelativistic kinematics):
\begin{equation}
S_{\alpha}^{\alpha'} = 1 - 2\pi i 
\sqrt{\mu_{\alpha}\mu_{\alpha'}k_{\alpha}k_{\alpha'}} \, 
T_{\alpha}^{\alpha'}(E+i0^{+};k_{\alpha'},k_{\alpha}) \,,
\end{equation}
with $k_{\alpha}$ the on-shell momentum for channel $\alpha$.

Our aim is to explore the existence of states above and below thresholds within the same formalism. Thus, we have to analytically continue all the potentials and kernels for complex momenta in order to find the poles of the $T$-matrix in any possible Riemann sheet.

\subsection{Line shapes calculation}

In order to describe the line shapes of the $Z_{cs}(3985)^-$ structure we follow the same procedure as in Ref.~\cite{Ortega:2018cnm}, which is briefly described here for the experimental analysis of BESIII~\cite{Ablikim:2020hsk}, though the same formalism can be used for the analysis of the LHCb Collaboration~\cite{Aaij:2021ivw}. The $Z_{cs}(3985)^-$ has been spotted in the $K^+$ recoil-mass spectrum of the $e^+e^-\to K^+(D_s^-D^{*\,0}+D_s^{*\,-}D^0)$ process at center of mass energy ranging from $4.628$ to $4.698$ GeV~\cite{Ablikim:2020hsk}.  The line shape of a $e^+e^-\to K^+Z_{cs}$ reaction from a point-like vertex at a given $\sqrt{s}$ and the subsequent $Z_{cs}\to AB$ is expressed as 

\begin{equation}
\frac{d\Gamma_{Z_{cs}\to AB}}{dm_{AB}} = \frac{1}{(2\pi)^3}\frac{k_{AB} k_{K Z_{cs}}}{4\,s}|{\cal M}^\beta(m_{AB})|^2  \,,
\end{equation}
with $\beta$ the quantum numbers of the channel $AB$, $m_{AB}$ is the invariant mass of the $AB$ meson pair and where $k_{KZ_{cs}}$ and $k_{AB}$ are the on-shell momentum of the $KZ_{cs}$ and $AB$ pair, respectively, given by
\begin{eqnarray}
k_{K Z_{cs}} &=& \frac{\lambda^{1/2}(\sqrt{s},m_{AB},m_K)}{2\sqrt{s}},\label{ec:kmomZpi}\\
k_{AB} &=& \frac{\lambda^{1/2}(m_{AB},m_A,m_B)}{2m_{AB}},\label{ec:kmomDD}
\end{eqnarray}
where $\lambda(M,m_1,m_2)=[(M^2-m_+^2)(M^2-m_-^2)]$, with $m_\pm=m_1\pm m_2$.

The Lorentz-invariant production amplitude, ${\cal M}$, describes the $Z_{cs}\to AB$ production and can be written as
\begin{equation}\label{eq:amplitude}
\mathcal{M^\beta}(m_{AB})=\left({\cal A}^\beta e^{i\,\theta_\beta}-\sum_{\beta'}{\cal A}^{\beta'}e^{i\,\theta_{\beta'}} \int d^3p\frac{t^{\beta'\beta}(p,k^\beta,E)}{p^2/2\mu-E-i\,0}\right).
\end{equation}
where $A^\beta$ and $\theta_\beta$ are parameters that describe the production amplitude and phase of the $(AB)_\beta$ channel from the $e^+e^-$ vertex.

Experimentally, discrete events are measured in the $\sigma(e^+e^-\to K^+ Z_{cs}^-)\times \mathcal{B}(Z_{cs}^- \to D_s^-D^{*0}+D_s^{*-}D^0)$ process and, hence, to describe the data we need to add a normalization factor to translate the decay rate into events:
\begin{equation}
N(m_{AB}) = \mathcal{N}_{AB}\times \frac{d\Gamma_{Z_c\to AB}}{dm_{AB}} \,.
\end{equation}
This normalization encodes other relevant process details such as the value of $\sigma(e^+e^-\to  K^+ Z_{cs}^-)$.

The set of  $\{{\cal A}_{AB},\theta_{AB},\mathcal{N}_{AB}\}$ parameters (for $AB$ the channels involved in the calculation) are, then, fitted by means of a global $\chi^2$ function minimization procedure using the available experimental data on $D_s^-D^{*0}+D_s^{*-}D^0$:
\begin{equation}
\label{eq:chisquare}
\chi^2(\{{\cal A,\theta,N}\}) = \sum_i \bigg(\frac{N^{\rm the}(x_i)-N^{\rm exp}(x_i)}{\sigma_i^{\rm exp}}\bigg)^2 \,,
\end{equation}
where the uncertainty of the parameters $\{{\cal A,\theta,N}\}$ are estimated from the Hessian.

\section{Results}
\label{sec:results}

Following our experience with the $Z_c(3900)^\pm$, we perform a coupled channel calculation for the $I(J^{P})=\frac{1}{2}(1^{+})$ for quark sector. We include those channels, in $^3S_1$ partial wave, whose thresholds are close to the experimental mass of the $Z_{cs}(3985)^-$. Those are the $J/\psi K^-$ (3592 MeV/$c^2$), $\eta_c K^{*\,-}$ (3877 MeV/$c^2$),   $D^-_s D^{*0}$ (3976 MeV/$c^2$), $D^0 D_s^{*\,-}$ (3979 MeV/$c^2$), $J/\psi K^{*\,-}$ (3990 MeV/$c^2$) and $D^{*0} D_s^{*-}$ (4120 MeV/$c^2$) channels, where the threshold energies are shown in parenthesis. 

We make two different calculation: One without annihilation diagrams (model $a$) an another including annihilation diagrams (model $b$). The reason for these two alternative approaches is that, without annihilation diagrams, non-diagonal $D^{(*)}D_s^{(*)}$ are decoupled and by comparing the two calculations one can evaluate the importance of these couplings.

\begin{figure}[th]
 \includegraphics[width=\columnwidth]{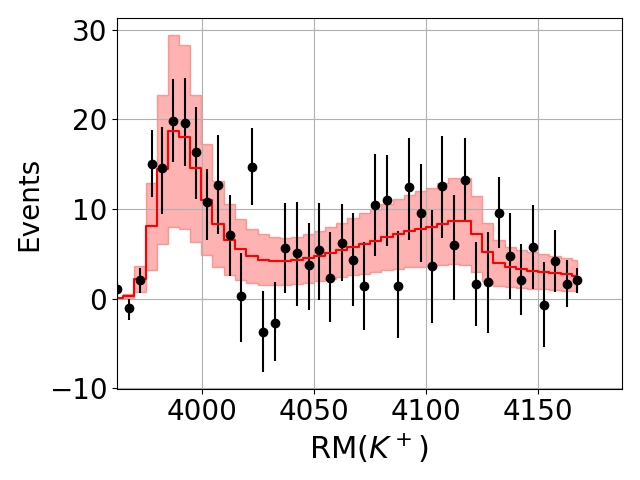} 
 \includegraphics[width=\columnwidth]{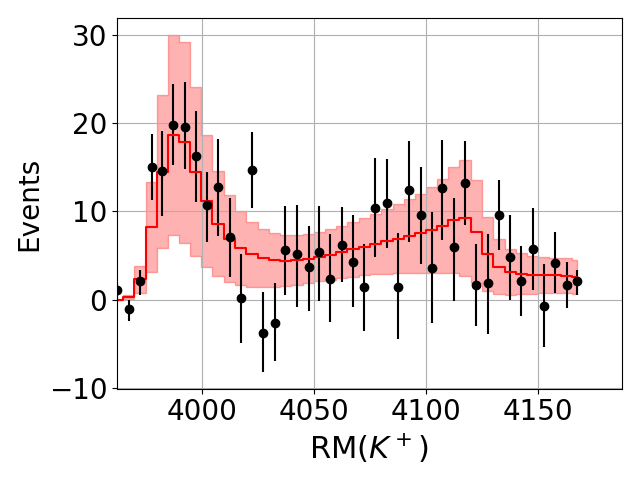}
 \caption{\label{fig1} Theoretical description (solid lines) of the experimental $K^+$ recoil-mass spectra (black dots) measured by BESIII~\cite{Ablikim:2020hsk}. The red shadowed-area around the line represents the 68\% CL of the fit. The upper pannel shows the calculation for {\it model a} and the lower pannel for {\it model b}. We remark here that the fit only affects the production part from the $e^+e^-$ vertex, with no fine-tuning of the CQM parameters in the description of the coupled-channels S-matrix.}
\end{figure}

From the analysis of the S-matrix we find two poles below the $D^-_s D^{*0}$ and $D^{*0} D_s^{*-}$ thresholds in the second Riemann sheet at $3970$ MeV/$c^2$ and $4110$ MeV/$c^2$ for the model $a$ and  $(3961-3\it i)$ MeV/$c^2$ and $(4106 - 5\it i)$ MeV/$c^2$ for the model $b$. The first one would be responsible for the $Z_{cs}(3985)^-$ peak, seen as an enhancement above the $D^-_s D^{*0}+D^0 D_s^{*\,-}$ thresholds, whereas the second one would be an unseen $Z_{cs}^-$ state, denoted as $Z_{cs}(4110)^-$, analog of the $Z_c(4020)^\pm$ in the charm-strange sector, which could give rise to the recent $Z_{cs}(4220)^+$ structure discovered by LHCb~\cite{Aaij:2021ivw}.

In our model,  both the $Z_{cs}(3985)^-$ and the predicted $Z_{cs}(4110)^-$ are not resonances but virtual state and, therefore, a direct comparison with the complex energy of the pole of a Breit-Wigner parametrization of a resonance would be misleading. That is the reason why the best way to confront our results with the experimental data is through the description of the line shapes.

Our results for the $K^+$ recoil-mass spectra are shown in the upper (model $a$) and lower panel (model $b$) of Fig.~\ref{fig1}.  The shaded area around the theoretical curve shows the statistical 68\%-confident level (CL) of the fit, obtained by propagating the errors of the fitted parameters by means of the covariance matrix. The values for the normalization factors and amplitudes are shown in Table~\ref{tab:norm}, the result on the $\chi^2/{\rm d.o.f.}$ is also collected therein. In order to describe the experimental measurement, the theoretical line shapes have been convoluted with the detector resolution.

\begin{table}
\begin{center}
 \begin{tabular}{c|cc|cc}
 \hline\hline
  & \multicolumn{2}{|c|}{BESIII data} & \multicolumn{2}{|c}{LHCb data} \\ 
  Parameters & Model a & Model b & Model a & Model b \\
  \hline
$\chi^2/{\rm d.o.f.}$  	&	1.00	&	1.02	&	2.65	&	2.04	\\
$\ln\left({\cal N}_{D_sD^*+DD_s^*}\right)$&	25.4(5)	&	24.6(6)	&	-	&	-	\\
$\ln\left({\cal N}_{J/\psi K}\right)$ 	&	-	&	-	&	25.22(6)&	25.43(8)\\
${\cal A}_{J/\psi K}$ 	&	0.71(4)	&	1.0(9)	&	0.026(1)&	0.028(2)\\
${\cal A}_{\eta_c K^*}$ &	0.31(3)	&	0.33(2)	&	0.39(2)	&	0.35(3)	\\
${\cal A}_{D_sD^*}$ 	&	0.028(2)&	0.052(5)&	0.140(4)&	0.02(1)	\\
${\cal A}_{DD_s^*}$ 	&	0.030(2)&	0.04(1)	&	0.00(2)	&	0.10(1)	\\
${\cal A}_{J/\psi K^*}$ &	0.01(1)	&	0.01(2)	&	0.9(1)	&	0.52(7)	\\
${\cal A}_{D^*D_s^*}$ 	&	0.17(5)	&	0.29(2)	&	0.143(4)&	0.15(1)	\\
$\theta_{J/\psi K}$ 	&    -1.42(3)	&	-2.65(8)&	-2.43(2)&	-0.39(13)\\
$\theta_{\eta_c K^*}$ 	&    -0.53(5)	&	-1.8(2)	&	1.43(4)	&	-3.19(6)\\
$\theta_{D_sD^*}$ 	&    -2.33(7)	&	3.2(4)	&	-3.19(3)&	-1.25(3)\\
$\theta_{DD_s^*}$ 	&    -2.39(7)	&	2.4(2)	&	1(6)	&	-0.96(11)\\
$\theta_{J/\psi K^*}$ 	&   -1.1(9)	&	-3(5)	&	-0.4(1)	&	1.15(13)\\
$\theta_{D^*D_s^*}$ 	&   -2.5(3)	&	-2.7(2)	&	0.67(2)	&	2.80(11)\\
  \hline\hline
 \end{tabular}
\caption{\label{tab:norm} Normalization and amplitude factors for the $D^-_s D^{*0}+D_s^{*-}D^0$ (left) and $J/\psi K^-$ (right) line shapes, which are fitted using Eq.~(\ref{eq:chisquare}). The minimum value of the $\chi^2/{\rm d.o.f.}$, calculated in the $[3.9,4.2]$ GeV energy range, is also given. The 68\% uncertainty in the parameters, in parenthesis, is obtained from the fit.}
\end {center}
\end{table}

One can deduced from these pictures that both model $a$ and $b$ provide a similar accuracy fit and, thus, the coupling among $D_s^{(*)}D^{(*)}$ does not appear to be relevant to reproduce the experimental data. The two peaks showed by the theoretical calculation correspond with the two virtual states we mentioned before. Then, our results reproduce the experimental data  without any fine-tuning of the model parameters used to describe the $Z_c(3900)^\pm$ and the $Z_c(4020)^\pm$, besides the unavoidable normalization and amplitude factors that describe the inner details of the production vertex, which involves further dynamics not relevant here. 

\begin{figure}[th]
 \includegraphics[width=\columnwidth]{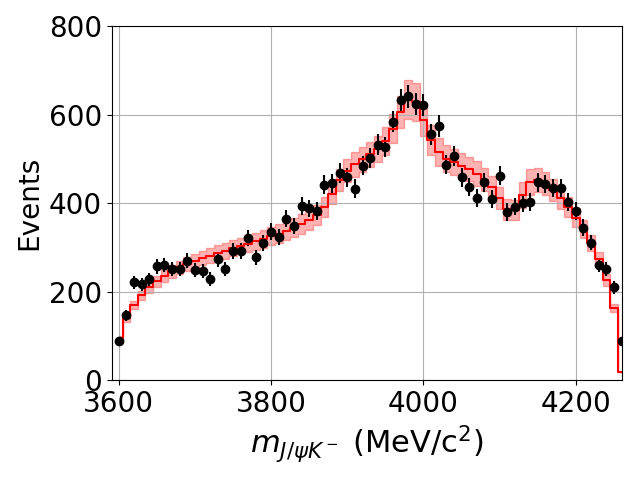} 
 \includegraphics[width=\columnwidth]{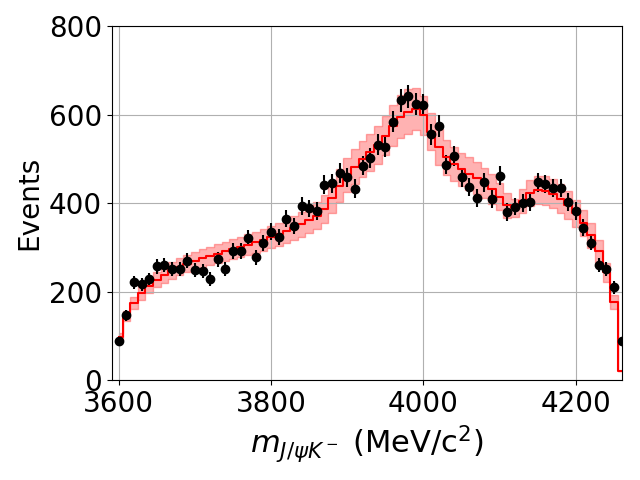}
 \caption{\label{fig2} Theoretical description (solid) of the experimental $J/\psi K^-$ invariant mass spectrum (black dots) measured by LHCb~\cite{Aaij:2021ivw}. Same legend as in Fig.~\ref{fig1}.}
\end{figure}

The same structures are able to reproduce the LHCb data on $J/\psi K^-$ invariant mass spectrum. As the reaction involved in BESIII and LHCb are of different nature ($e^+e^-$ vs $pp$ collisions) and the center of mass energy is not the same, we do not expect that the same parameters that reproduce the $K^+$ recoil-mass events from BESIII hold in the $J/\psi K^-$ data from LHCb. That is why we have performed a second fit on LHCb data, whose results are shown in the upper (model $a$) and lower panel (model $b$) of Fig.~\ref{fig2}. Both models are able to reproduce the experimental data, with equivalent accuracy, but model $b$ gives a slightly better $\chi^2/{\rm d.o.f.}$, shown in Table~\ref{tab:norm}. It is worth noticing that, besides the different production weights involved in both reactions, which can be different, it is the pole structure of the S-matrix that gives the corresponding peaks that appear in the experimental data. Thus, our results support the hypothesis that the $Z_{cs}(3985)^-$ and the $Z_{cs}(4000)^+$ are the same state, and that the $Z_{cs}(4220)^+$ is an effect of an event dip around the $D^{*0} D_s^{*-}$ threshold, which emerges as a second peak in the $D^-_s D^{*0}+D_s^{*-}D^0$ line shape, around the same energy.

In addition, we can exploit the same framework to predict states in the hidden bottom strange sector within the same procedure we have used in this calculation, except the obvious change of the charm quark mass for the bottom quark mass. The main channels involved in the calculation are $B^{*-} B^0_s$, $B^{-} B^{*0}_s$ and $B^{*-} B^{*0}_s$~\footnote{The analogous hidden-bottom channels $\Upsilon(1S)K^{(*\,-)}$ and $\eta_b(1S)K^{*\,-}$ are too far away to affect the pole determination, though they are relevant to describe the lineshapes and decays}.  This time we obtain two poles in the second Riemann sheet below the $B^{*\,-}B_s^0$ and $B^{*\,-}B_s^{*\,0}$ thresholds, corresponding to two virtual $Z_{bs}$ states at $10691$ MeV/$c^2$ and $10739$ MeV/$c^2$, respectively. These states should be identified as the $SU(3)_F$ partners of the $Z_b(10610)^\pm$ and the $Z_b(10650)^\pm$, heavy partners of the $Z_{cs}(3985)^-$ and $Z_{cs}(4110)^-$, and could be detected in the $\Upsilon(1S)K^-$ and $B^{*-} B^0_s+B^{-} B^{*0}_s$ channels.


\section{Summary}
\label{sec:summary}
The $Z_{cs}(3985)^-$ is a new signal which confirms that we need to go beyond the naive $q\bar q$ structures to describe meson states and consider more complex structures. The proximity of the $D_s^{(*)}D^{(*)}$ thresholds suggest that this state may be the corresponding partner of the $Z_c(3900)^\pm$ in a $SU(3)_F$ scheme.

To explore its structure, we have performed a coupled-channels calculation for the $I(J^{P})=\frac{1}{2}(1^{+})$ four-quark sector in the framework of the chiral constituent quark model. We included the most relevant meson-meson channels with mass thresholds close to the experimental $Z_{cs}(3985)^-$ mass. The calculation is done using the standard set of parameters of Ref.~\cite{Vijande:2004he}, which also reproduced the $Z_c(3900)^\pm$ and $Z_c(4020)^\pm$ line shapes~\cite{Ortega:2018cnm}. In that sense, the calculation of the poles is parameter-free.

The experimental $K^+$ recoil-mass spectra is well reproduced whether the full coupling with $D_s^{(*)}D^{(*)}$ is included or not through annihilation interactions, pointing to the importance to couple with hidden-charm channels such as $J/\psi K^{(*)}$ and $\eta_c K^*$, as it was the case for the original calculation of $Z_c(3900)^\pm$~\cite{Ortega:2018cnm}. In this respect, the dynamics involved in the description of the $Z_{cs}(3985)^-$ is similar to the one of the $Z_c(3900)^\pm$, but in the latter pion-exchange interactions were important, which we do not have in this sector. The analysis of the S-matrix poles allows us to conclude that the structures of the line shape emerge due to the presence of two virtual states that produce two distint cusps at the $D_s^-D^{*\,0}+D_s^{*\,-}D^0$ and the 
$D_s^{*\,-}D^{*\,0}$ thresholds. Moreover, our calculation predicts a new $SU(3)_F$ partner of the $Z_c(4020)^\pm$, called $Z_{cs}(4110)^-$. 
These results support  the line shapes fit of Ref.~\cite{Wang:2020htx} and the conclusion of Ref.~\cite{Yang:2020nrt} about the virtual nature of the $Z_{cs}(3985)^-$ structure.
Besides, we are able to reproduce the recent LHCb data~\cite{Aaij:2021ivw}, reinforcing the hypothesis that the $Z_{cs}(3985)^-$ and the $Z_{cs}(4000)^+$ are the same state measured in different final channels.

As an additional result, we found two virtual states at $10691$ MeV/$c^2$ and $10739$ MeV/$c^2$ in the bottom-strange sector which could be the partners of the well established $Z_b(10610)^\pm$ and the $Z_b(10650)^\pm$ states.

Looking for these states together with further searches of the $Z_{cs}(4110)^-$, which could be the recent $Z_{cs}(4220)^+$ announced by LHCb, would be an interesting topic for future experiments.


\begin{acknowledgments}
This work has been funded by
Ministerio de Ciencia e Innovaci\'on
under Contract No. PID2019-105439GB-C22/AEI/10.13039/501100011033
and by EU Horizon 2020 research and innovation program, STRONG-2020 project, under grant agreement No 824093.
\end{acknowledgments}


\bibliography{paperZcs}

\end{document}